# Novelty in news search: a longitudinal study of the 2020 US elections

Ulloa, Roberto[1(*)]   Makhortykh, Mykola[2]   Urman, Aleksandra[3]   Kulshrestha, Juhi[4,5]

[1]GESIS – Leibniz Institute for the Social Sciences

[2] University of Bern

[3]University of Zurich

[4]Aalto University

[5]Universty of Konstanz

(*) Corresponding Author: Ulloa, Roberto (roberto.ulloa@gesis.org)

## Authors Note

**Disclosure Statement:** Nothing to declare.

**Acknowledgments:** Funded by the Deutsche Forschungsgemeinschaft (DFG, German Research Foundation) - Projektnumber 491156185. Data collections were sponsored from the SNF (100001CL_182630/1) and DFG (MA 2244/9-1) grant for the project "Reciprocal relations between populist radical-right attitudes and political information behaviour: A longitudinal study of attitude development in high-choice information environments" led by Silke Adam (U of Bern) and Michaela Maier (U of Koblenz-Landau) and FKMB (the Friends of the Institute of Communication and Media Science at the University of Bern) grant "Algorithmic curation of (political) information and its biases" awarded to Mykola Makhortykh and Aleksandra Urman.

# Abstract

The 2020 US elections news coverage was extensive, with new pieces of information generated rapidly. This evolving scenario presented an opportunity to study the performance of search engines in a context in which they had to quickly process information as it was published. We analyze novelty, a measurement of new items that emerge in the top news search results, to compare the coverage and visibility of different topics. We conduct a longitudinal study of news results of five search engines collected in short-bursts (every 21 minutes) from two regions (Oregon, US and Frankfurt, Germany), starting on election day and lasting until one day after the announcement of Biden as the winner. We find more new items emerging for election related queries ("joe biden", "donald trump" and "us elections") compared to topical (e.g., "coronavirus") or stable (e.g., "holocaust") queries. We demonstrate differences across search engines and regions over time, and we highlight imbalances between candidate queries. When it comes to news search, search engines are responsible for such imbalances, either due to their algorithms or the set of news sources they rely on. We argue that such imbalances affect the visibility of political candidates in news searches during electoral periods.

# Introduction

The 2020 US elections were one of the most viewed events of 2020, attracting 56.9M viewers on cable and broadcast TV at prime time alone (Nielsen, 2020). As shown by the record turn-out (Schaul et al., 2020), the stakes were high in a polarized nation (Boxell et al., 2020) whose citizens were deciding the direction of a major international power. Media outlets were ready to cover every detail that would keep their visitors engaged, reporting novel pieces of information every few minutes, if not seconds (e.g., Astor, 2020; Kommenda et al., 2020). At a proportional pace, digital intermediaries, such as search engines, frantically processed the material to show the latest and most relevant updates to their audience. This scenario presented an opportunity to explore the performance of search engines under an intensively mediated political campaign in which political actors competed for the spotlight (Kaid & Strömbäck, 2008). This paper reports how search engines covered the elections in terms of novelty, i.e., inclusion of novel items among their top news results, which, we argue, is essential for analyzing the coverage that a topic receives by the search engine.

Earlier research has shown that success in elections depends on the attention that the media spends on candidates (Hopmann et al., 2010; Maddens et al., 2006; Reuning & Dietrich, 2019; van Erkel et al., 2020). Coverage (and visibility) has not directly been addressed in news search scholarship because of the reactivity of search engines, namely search engines do not feature a selection of materials per se (e.g., as in a news website), but retrieve them in response to user queries. For any query, search engines return a long list of news articles, albeit in the majority of cases individuals will interact with only those at the top (Pan et al., 2007; Urman & Makhortykh, 2021). Because the relevance of news items changes over time, more relevant items can appear at the top when the individual searches again. This has consequences for the visibility of the topic, as the individual would be exposed to a more diverse set of news when the novelty is higher.

In this paper, we analyze news search by investigating the novelty of results that emerges for 9 queries: 3 related to the US elections ("joe biden", "donald trump" and "us elections"), 3 topical ("coronavirus", "poland abortion", "nagorno-karabakh conflict") and 3 stable ones ("first world war", "holocaust", "virtual reality"). Data were obtained during the 2020 US presidential elections from 5 search engines (Google, Bing, DuckDuckGo, Yahoo! and Baidu); snapshots for each query were captured every ~21 minutes between Nov 3rd, 07:31 a.m. and Nov 9th, 06:40 a.m. Eastern Time (ET) using 240 virtual agents located in two geographical areas: Oregon (United States) and a non-US location Frankfurt (Germany). Our focus is to investigate the evolution of news results across four periods, defined by three key events: (a) close of all polls, (b) call of Michigan's results, the 45th state being called followed by 3 days without any calls, and (c) call of Pensylvannia's results, the state that indicated the victory to Biden.

Following (Kulshrestha et al., 2019), we include weights corresponding to the search ranking in our novelty metric to capture the tendency of individuals to click on top results more often (Pan et al., 2007; Urman & Makhortykh, 2021), including news articles (Ulloa & Kacperski, 2022). To analyze the data, we use linear mixed models, with repeated measures that stem from our longitudinal observations. First, we present evidence that novelty is indeed higher for election related queries, as well as for the COVID-19 pandemic, but neither for localized happenings outside of the US (e.g., Poland abortion protests), nor for stable queries. Then, we demonstrate that the novelty for the two candidates differs across search engines and regions, and that the novelty is disproportionally high, in particular for the query "donald trump" in Bing and Oregon.

## Media coverage and elections

Neither voters' positions on political issues nor the candidates' personal traits matter if the candidate is not visible to the voter (Hopmann et al., 2010). Previous research shows that electoral success depends on the attention that the media pay to candidates (e.g., Hopmann et al., 2010; Maddens et al., 2006; Reuning & Dietrich, 2019; van Erkel et al., 2020). For example, observers have attributed Donald Trump's victory in 2016 to the amount of news coverage he received compared with his rivals (Shafer, 2016). Although news reports are guided by journalistic norms (Hackett, 1984; Muñoz-Torres, 2012), research indicates that there are market forces that influence the gatekeeping aspect of the media (Hamilton, 2011; Patterson, 2013), and that these factors were exploited by Trump during the 2016 elections (Callum Borchers, 2016; Confessore & Yourish, 2016).

Last-minute broadcasts which inform viewers about elections are of particular interest for the discussion of factors affecting electoral choices (Hofstetter & Buss, 1980). Such information is relevant for late deciding voters, the numbers of which have been rising in Western democracies, including in the US (see Yarchi et al. (2021) for a list of countries). For example, on election day, 12.5% of 2016 US voters were either undecided or said they planned to vote for third-party candidates (Silver, 2017). Not surprisingly, late deciding voters sometimes determine the final outcome of elections (Box-Steffensmeier et al., 2015; Schill & Kirk, 2017; Schmitt-Beck & Partheymüller, 2012). Voters that remain undecided are considered very unpredictable (Box-Steffensmeier et al., 2015, p.; Gopoian & Hadjiharalambous, 1994); they appear more reactive to campaign coverage (Fournier et al., 2004) and less critical about the information they consume (Samuel-Azran et al., 2022).

The period that follows the elections is also a sensitive one, as the legitimacy of the process is called into question by some elites that spread rumors of fraud (Minnite, 2011). Such rumors characterized the electoral campaign of 2020 US elections (Benkler et al., 2020; Berlinski et al., 2021; Enders et al., 2021) which were also accompanied by Trump's threats of not committing to a peaceful transfer of power (Crowley, 2020). Such claims continued after the election, including the period covered by our data collection[1], leading to Trump supporters storming the capitol on January 6th (CNN, 2021).  Hence, the post-election period is critical because the rumors are more likely to affect populations that are dissatisfied with the outcome.

---

[1] There are 98 fraud claims related to the elections attributed to Trump in The Washington Post database during our collection period (Nov 3rd to Nov 11th, 2020), 79 occurred after the election day https://www.washingtonpost.com/graphics/politics/trump-claims-database/

## Search engines as digital intermediaries

News organizations are becoming more dependent on digital intermediaries, such as search engines and social media platforms. These intermediaries represent short-term opportunities to engage audiences, even if these opportunities might result in the loss of control over their organization professional identity (Nielsen & Ganter, 2018). The technological companies behind these intermediaries are also leveraging their role to shape political communication (Kreiss & Mcgregor, 2018), while parties and candidates try to adapt their campaigns to the new media logic (Klinger & Svensson, 2015).

We focus on search engines, as they play a gatekeeper role in the current high-choice information environments (Van Aelst et al., 2017). Individuals frequently use them to seek information (Urman & Makhortykh, 2021) and learn from the results obtained (Fisher et al., 2015; Ward, 2021). Moreover, individuals rely on search engine ranking algorithms as a measure for content relevance (Edelman, 2021; Keane et al., 2008; Schultheiß et al., 2018; Urman & Makhortykh, 2021). Consequently, search engines became one of the most used technologies of finding political information (Dutton et al., 2017), which is crucial as there is evidence of their potential to shift voting preferences of undecided voters (Epstein et al., 2017; Zweig, 2017).

Specifically, we are interested in the coverage of topics in search engines. Instead of looking at a single result page in which virtually all items presented are pertinent to the query, we look at novelty, i.e., the number of novel items that emerge in the top results. We argue that higher novelty increases the visibility of the topic. First, an individual is more likely to encounter more information if they search more than once for the same topic at different points in time. Second, it increases the potential amount of information that can be circulated via the searcher's personal network due to the effects of interpersonal communication (Katz & Lazarsfeld, 2017; Schmitt-Beck, 2003). Third, given that recency plays an important role in the ranking of results (Dong et al., 2010), there could be spillover effects to other elements of search engine interfaces (e.g. news featured in the main search results).

## Search engine auditing

Search engines have attracted a lot of attention in the algorithm auditing field, which investigates performance of algorithmic systems and their potential biases (Mittelstadt, 2016). First, researchers have reported a concentration of results of a few news sources for different Google interface components such as the main search results (Jiang, 2014), Google Top Stories (Kawakami et al., 2020; Trielli & Diakopoulos, 2019), news search (Nechushtai & Lewis, 2019) and video search (Urman et al., 2021a). These findings extend to the Dutch (Courtois et al., 2018) and German context (Haim et al., 2018; Unkel & Haim, 2019).

Second, Pariser (2011) argued that search personalization, i.e., content selected according to previous individual's consumption and preferences, could lead to filter bubbles, i.e., feedback loops of information which hinder exposure to different views. Current empirical evidence indicates that such concerns are overstated, and that, instead, search engines can lead to an increase of diversity of news sources that are consumed (e.g., Stier et al., 2022; Ulloa & Kacperski, 2022).

Third, several aspects of political representation have been investigated. Puschmann (2019) finds that some political parties and candidates can exert greater influence over how they are represented in search media (in terms of source type) than others. There is also evidence suggesting a (modest) left partisan leaning in Google search results (Robertson et al., 2018; Trielli & Diakopoulos, 2019), although the leaning is usually measured on the source and not necessarily the content level (Ganguly et al., 2020).

Only few studies conduct longitudinal investigations: Metaxas and Pruksachatkun (2017) reported that Google (but not Yahoo! and Bing) restricted variation of sources across time, favoring those that were considered "reliable" to prevent the surfacing of "fake news". Kawakami et al. (2020) found that a year before the US elections 2020, the number of unique news in Google's Tops Stories differed for different candidates, and it was higher for Donald Trump, which was attributed to him being the incumbent president. Pradel (2021) found gender and party differences in the amount of personal information related to politicians that appears on the search suggestions before and after the elections. Closer to our work, Metaxa et al. (2019) systematically analyzed daily search results, finding search outputs to be relatively stable, though some shifts suggested the existence of internal algorithmic factors, e.g., monthly synchronization of Google servers.

Most of the works have investigated Google exclusively, however there are exceptions that demonstrate differences between search engines in terms of source concentration (Jiang, 2014), "gaming" or "link bombing" during the 2008 US Congressional Elections (Metaxas & Mustafaraj, 2009), content diversity (Steiner et al., 2020), preventing "fake news" (Metaxas & Pruksachatkun, 2017) and results overlap (Urman et al., 2021b).

## Research questions and hypothesis

Our aim is to analyze the rate at which new information is incorporated in the search results of different queries, search engines, and regions. To our knowledge, this is the first time that novelty of news search results is analyzed, i.e., we give first insight into the pace at which information is integrated into the search engines. The fine granularity of our data collection (every 21 minutes per query) allows us to capture sudden changes.

We first contrast US-related queries with other topical queries – we chose COVID-19 ("coronavirus"), the Poland abortion protests following the Constitutional Tribunal ruling on October 22, 2020 ("poland abortion") and the 2020 Nagorno-Karabakh conflict dated 27 September 2020 – 10 November 2020 ("nagorno-karabakh conflict"), for which we also expected relatively high coverage and novelty of news articles. Additionally, we included stable queries ("first world war", "holocaust", "virtual reality"), for which we expected a low amount of novel news. These categories serve as a benchmarks to demonstrate the coverage given to novel items related to the US elections, see RQ1 in Table 1.

We further examine the evolution of the novelty for the US elections related queries. First, we divide our collection in four periods (denoted with roman numbers: I, II, III, and IV) defined by three key events: (a) close of all polls (Nov $4^{th}$, 1:00 a.m. ET), (b) call of Michigan's result (Nov $4^{th}$, 5:58 p.m. ET) , the $45^{th}$ state being called followed by 3 days without any calls, and (c) call of Pennsylvania's results (Nov $7^{th}$, 11:25 a.m. ET), the state that gave the final victory to Biden. Then, we examine differences between periods, regions, and search engines. We pay special attention to differences between the queries of the two candidates to find imbalances in novelty. See RQ2 in Table 1.

| ID | Research question and hypotheses | Supported |
|---|---|---|
| **RQ1** | ***Is the novelty of queries related to the US elections higher than other queries during election day and the hours following it (*)?*** | |
| H1a | The novelty for queries related to the US elections is higher than for queries related to other topics, especially those not news-worthy during the same period (i.e., stable queries such as the First World War). | **Consistently** |
| H1b | More novelty is displayed for topical queries during the collection (e.g., "poland abortion") than those not news-worthy (e.g., "first world war"). | **Partially** |
| **RQ2** | ***Are there differences in novelty for the different US elections related queries, regions, periods and search engines?*** | |
| H2a | More novelty is displayed in Oregon (United States) than in Frankfurt (Germany). Given the role that localization plays in search results (Kliman-Silver et al., 2015), we assume that more attention is drawn to the topic in the US. | **Consistently** |
| H2b | There are differences in the novelty of results shown by different search engines. | **Consistently** |
| H2c | Specifically, Google will display less novelty than Bing and Yahoo! as previous research indicates that their organic results show less variation over time, presumably as a consequence of potential mechanisms to control web spammers (Metaxas & Pruksachatkun, 2017). We assume a similar trend for news results. | **Partially** |
| H2d | Novelty of the US queries diminishes as results are more distant from election day and stories become less abundant in news. | **Consistently** |
| H2e | There are no differences between novelty of the candidates in different search engines before the announced election result (Periods I, II, III). | **Rejected** |
| H2f | Novelty of results for "joe biden" will be higher than for "donald trump" after the declaration of Biden as a winner (Period IV). | **Consistently, but not for all period IV** |

**Table 1. Research questions and related hypotheses of the present study.** The first column identifies the research question or hypothesis presented in the second column. The third column indicates if the hypothesis is supported or not: consistently, means that almost all (or all) cases side with the hypothesis; partially, if there are notable counterexamples that need to be explained; and rejected, if most of the evidence sides with the opposite direction of the hypothesis. (*) Specifically for RQ1, from Nov 3rd, 07:31 a.m. to Nov 4th, 11:10 a.m. ET.

# Materials and Methods

For our data collection, we used virtual agents, i.e., software that simulates human behavior (Ulloa et al., 2021). The implementation of such an agent took the form of a browser extension (for Firefox and Chrome) that simulates the navigation of search result pages on a search engine, and that collects the HTML of the pages by sending it to a server. The agent collects at least 50 news search results (if available), and it iterates over the list of terms until terminated. Before starting the search for a new query, the browser data (e.g., history, cache) is cleaned, thus avoiding personalization effects based on previous browsing history. We parsed the HTML pages to extract the top organic news results of each search routine.

## *Data collection*

We used the news search engine results collected in two consecutives experiments which included 9 terms divided equally in three categories (see Table 2). A category was assigned to each agent, and each of the three terms in the category were queried sequentially in a continuous loop, so that each term was searched every 21 minutes (a search routine lasts 7 minutes per term). The data was collected from Nov 3rd, 07:31 a.m. to Nov 4th, 11:10 a.m. ET, accounting for 80 rounds (collection A). Additionally, the collection for the US category (US-related queries) was extended until Nov 9th, 06:40 a.m. ET (extra 329 rounds, collection B).

| Query Category | Terms | Related to | Collections |
|---|---|---|---|
| US | joe biden, donald trump, us elections | US Elections 2020 | A, B |
| topical | coronavirus, poland abortion, nagorno-karabakh conflict | Issues that were highly covered by the news at the time | A |
| stable | first world war, holocaust, virtual reality | Topics that we consider were not being of news importance at the time | A |

**Table 2. Terms of each query category.** The first column displays the name of the query category, the second column the terms included in the category, the third column the topic they are related to, and the fourth column, the experiment (s) in which they were included

For collection A, a total of 240 virtual agents were deployed simultaneously in the Amazon Elastic Compute Cloud (using 120 CentOS virtual machines, each hosting two virtual agents: one in Chrome and one in Firefox), and the agents were distributed equally to each experimental condition given by the combination of variables in Table 3. In total, each experimental condition was assigned to 4 different agents, so that we could account for the effects of results' randomizations by the search engines (Makhortykh et al., 2020). Additionally, all machines on a given region were allocated in the same range of Internet protocol (IPs). For collection B, we reduced the scale of the experiment to keep costs under our budget (as an election winner did not emerge until days after), so all machines assigned to the topical and stable categories were terminated and only 1 agent per condition was kept for the US category (20 agents in total).

| Variable | Values | N |
|---|---|---|
| Region | Oregon, Frankfurt | 2 |
| Browser | Chrome, Firefox | 2 |
| Search Engines | Baidu, Bing, DuckDuckGo, Google, Yahoo! | 5 |
| Query Categories | US Elections, topical, stable | 3 |

**Table 3. Variables of the experiment.** The first column displays the name of the factor, the second column the possible values of each factor, and the third column the number of values in each variable.

In Appendix S1, we include a detailed analysis of the data collection coverage. In general, very good coverage can be reported for our analyses and although some systematic issues are reported, we make sure that our analysis are not directly affected by them. Additionally, the weighting of the ranking, presented in the next section, improve on potential distortions.

## Definitions and metrics

**Item.** It describes the combination of a URL and a title in a news search result. An item is the main unit of analysis in this paper because some URLs are used as live streams (e.g., https://www.nytimes.com/live/2020/11/07/us/biden-trump) to dynamically publish different pieces of information. Thus, the URL does not uniquely identify a news search result.

**New items.** We define that an item in round *j* is new if it is the first time that it appears for a given query term and virtual agent; conversely, an item is not new if it appears in any previous round *i* (i.e., *i < j*) for that term and agent. The following items are discarded as we cannot ascertain if they are new or not: (1) items that appear on the first (successful) collection round, (2) items of a round *j* that follows a missing or incomplete round *j-1*.

**Weighted rank.** All our metrics (except diversity) consider the position (rank) of the search results. For this, we generalize the weights used to estimate the (political) biases on search results (Kulshrestha et al., 2019). In their work, each rank in the list is assigned a weight such that higher weights are assigned to higher ranked results (i.e., top results), which is then multiplied by the (political) bias score of the corresponding item. Let $L$ be the sequence of items ($i$) of size *N* corresponding to the top results of a query in a given round, the weight for the rank $r$ is calculated as follows:

$$W(r, N) = \frac{1}{N} \sum_{c=r}^{N} \frac{1}{c}$$

**Novelty.** We define a parameter $\delta$ that takes the values 1 or 0 ($\delta$) depending on whether the item is new or not, and use the weighted rank measure to calculate the novelty of the sequence $L$:

$$Novelty(L) = \sum_{i=1}^{|L|} \delta_i \cdot W'(i, |L|)$$

where $W'$ is a re-scaled weight that accounts for missing items, otherwise they would be implicitly counted as zeros, i.e., not new news items. We assume that missing items of an incomplete round should occur independently, therefore counting them as 0 would bias the calculation (decreasing the novelty). Let $L'$ be the set of collected items, then the weights are re-scaled as follows:

$$W'(r, N) = \frac{W(r, N)}{\sum_{i=1, l_i \in L'}^{N} W(i, N)}$$

Note that the novelty ranges from 0 to 1. To give an intuitive idea, a new item that appears of the top position (out of 50) represents a novelty of 0.089 (i.e., 8.9% change of the information assuming that the weights are an adequate way of representing the relevance of the results), whereas a change on the 10th represents a novelty of 3.3%. The novelty of a change in both, the 1st and 10th, results is represented by the addition of the two novelties, i.e., 12.3%.

Note that the novelty ranges from 0 to 1. To give an intuitive idea, a new item that appears at the top position (out of 50) represents a novelty of 0.089 (i.e., 8.9% change of the information assuming that the weights are an adequate way of representing the relevance of the results), whereas a change on the 10th represents a novelty of 0.033. The novelty of a change in both, the 1st and 10th position results is represented by the composition of the two novelties, i.e., 0.122%.

### *Study design and analysis*

Our study considers several factors that affect the search results: search engine, region, query (or query category) and period. The first three are described in Table 2 and Table 3. We define four periods according to three key events (close of all polls, call of Michigan's result and call of Pennsylvania's results). As an independent (continuous) variable, we analyze novelty as described before.

To answer the research questions (Table 1), we used linear mixed-effect models (Bates et al., 2015), fitting the interaction between the study factors (query or query category, period, engine and region). We considered the following random intercepts for repeated measures: query term (when query category is one of the factors), agent and round. However, we only report the models with the lowest Akaike's Information Criterion (AIC) (Akaike, 1974); in case of models not being statistically different, we kept the simplest of them. For novelty, we tested two types of models according to RQ1 (query category, engine, and region) and RQ2 (query, engine, region, period).

To evaluate our hypotheses, we count the relevant contrasts that are significantly different and support the hypothesis (or contradict it). The contrasts are calculated on the fitted model using the emmeans R package (Lenth, 2021). All our plots include bootstrapped confidence intervals (95%); in the case of time series, rolled averages (and confidence intervals) are calculated based on the observations of the previous 6 hours.

## Results

We found a triple interaction between the query category, engine, and region; $F(8, 207.266) = 9.205$, $p < .001$ (Appendix S4). The US-related queries displayed significantly more novelty than the topical and stable queries for Bing, DuckDuckGo and Google in both regions (.10 < $\beta$ < .23, $p < .007$) except between US- and topical-related queries for DuckDuckGo in Frankfurt. No significant differences were found between the topical and stable queries. Thus, we found support for H1a, but not for H1b. Figure 1 presents the results by query indicating that "coronavirus" is carrying the effect of the topical category. To confirm this, we fitted another model (Appendix S2) with an exclusive category for the "coronavirus" query, which was consequently removed from the topical category. In this new model, the US-related queries displayed significantly more novelty than the topical and stable queries for all regions and engines (.08 < $\beta$ < .22, $p < .001$), except for Baidu (NS). Additionally, for Google, the US-related queries displayed significantly more novelty than the "coronavirus" query (-.13 < β < -.07, p < .001). Given the generally low novelty of Baidu, we will not consider it for the rest of the analysis.

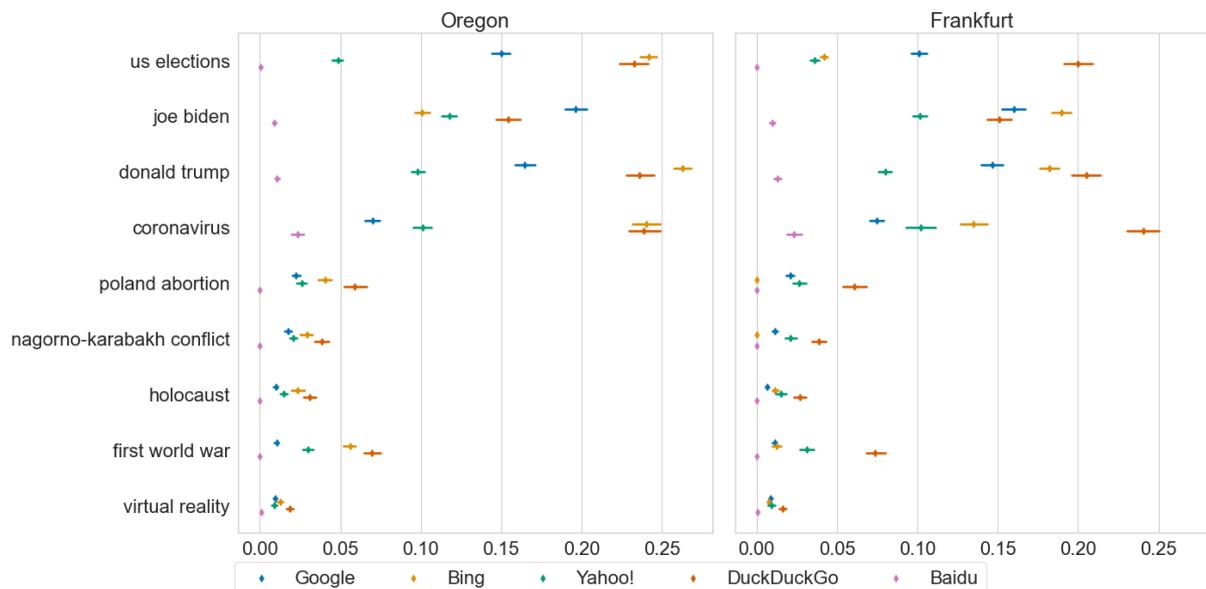

**Figure 1. Novelty of query terms.** The Y-axis shows the query terms that were explored, and the X-axis shows the novelty (truncated to .3, maximum theoretical value: 1.0). The legend shows the different search engines. The left plot corresponds to Oregon and the right plot to Frankfurt. Bootstrapped confidence intervals at 95%.

To analyze the difference between the US-related queries, we fitted a model (Appendix S6) including the three queries and the four periods (Figure 2). We found a quadruple interaction; F(18, 789420.858) = 17.45, p < .001. To understand the patterns of this interaction we analyzed the contrasts in four steps according to our hypotheses. First, we analyzed the hypothesis that the novelty was higher for Oregon (H2a), which was supported by 9 (out of 16) contrasts for "us elections" (-.11 < β < -.03, p < .001), by 6 (out of 16) for "donald trump" (-.11 < β < -.03, p < .001) and 1 (out of 16) for "joe biden" (β = -.06, p < .001), and rejected by 4 (out of 16) contrasts for "joe biden" (all corresponding to Bing; .03 < β < -.15, p < .001), 1 (out of 16) contrast for "donald trump" (β = .037, p = .001) and none for the "us elections".

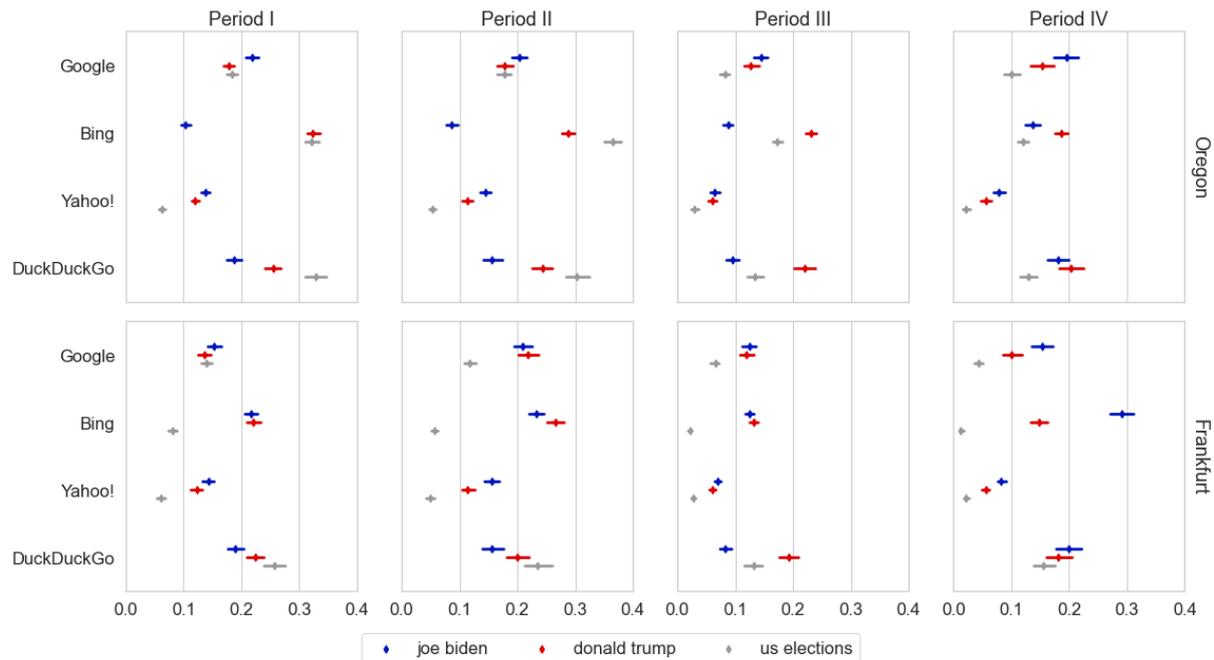

**Figure 2. Novelty of US-related queries across periods.** The X-axis shows the novelty truncated to .4 (maximum theoretical value: 1.0), and the Y-axis the engines. The legend identifies the US-related queries. The top row corresponds to Oregon and the bottom one to Frankfurt. The columns present the results per period. Confidence intervals at 95%.

Second, we found significant differences between the novelty of different search engines (H2b) as shown in Table 1. Yahoo! consistently displayed the least novelty, while DuckDuckGo always occupied the first or second position. Bing occupied the first position in 4 (out of 6) combinations of query regions but shared the last position with Yahoo! for "joe biden" in Oregon, and "us elections" in Frankfurt. Google occupied the third position 3 times, the second 2 times and the first one time. Therefore, we only find partial support for H2c: Google displayed more novelty than Yahoo! in all cases and less novelty than Bing in 4 out of 6 cases; this held true for all periods regardless of the changes observed in specific periods (see last column of the table).

| Region | Query | 1st | 2nd | 3rd | 4th | Periods |
|---|---|---|---|---|---|---|
| Oregon | joe biden | Google (.19) | DDG (.16) | | Yahoo! (.11), Bing (.10) | I, II, *III, IV* |
| | donald trump | Bing (.26) | DDG (.23) | Google (.26) | Yahoo! (.09) | I, II, III. IV |
| | us elections | Bing (.25) | DDG (.22) | Google (.14) | Yahoo! (.04) | *I*, II, III, *IV* |
| Frankfurt | joe biden | Bing (.21) | Google (.16), DDG (.16) | | Yahoo! (.12) | *I*, III, *IV* |
| | donald trump | DDG (.20), Bing (.19) | | Google (.14) | Yahoo! (.09) | I, III, IV |
| | us elections | DDG (.19) | Google (.09) | Bing (.04), Yahoo! (.04) | | I, II, *III, IV* |

**Table 4. Engines sorted according to novelty.** The first column displays the region, and the second the query. Column 3 to 6 indicates the position that the search engine took according to their novelty (in parenthesis); if there is no statistical difference between two engines, they are displayed in the same cell separated by column.

The last column indicates the periods for which the order held true; the italics indicates when the order of the non-statistical differences were switched.

Third, we analyzed the novelty of subsequent periods on all US-related queries for Oregon (H2d): 65 (out of 144) contrasts supported the hypothesized downward trend as time passed from election day (.04 < β < .21, p < .008). 10 contrasts contradicted the hypothesis (-.15 < β < -.04, p < .001), out of which, 6 involved period IV for "joe biden" which can be explained by the spike of news for "joe biden" after he was declared the winner (Period IV, Figure 3).

Fourth, we analyzed the contrasts between the candidates queries in Oregon for Period I to III (H2e). For Oregon, 8 out of 12 contrasts contradicted the hypothesis of unbiased novelty in Oregon; this included all contrasts of Bing (.14 < β < .22, p < .001), and DuckDuckGo (.06 < β < .13, p < .001), where we found more novelty for "donald trump" than for "joe biden", and one for Google (Period I, β = -.04, p < .001) and Yahoo! (Period II, β = -.03, p < .001), in which we found the opposite. The unbalance for Bing in Oregon is particularly disproportionate: at the end of the Period III, there are 3.24 times as many unique news items for "donald trump" (N=3599) as there are for "joe biden" (N=1110). This is followed by DuckDuckGo, with 1.99 times as many results for "donald trump" in Oregon (and 1.77 in Frankfurt, see Appendix S7 for other proportions). For Frankfurt, the results were more balanced: only 3 out of 12 contradicted the hypothesis in the same directions, according to search engine: Bing (Period II, β =.03, p < .001), DuckDuckGo (Period III, β = -.11, p < .001) and Google (Period II, β = -.04, p < .001).

Fifth, we analyzed novelty displayed by the candidate queries in Period IV (H2f): 3 (out of 4) contrasts in Frankfurt supported the hypothesis that Biden's query would display more novelty than Trump's (-.15 < β < -.02, p < .003). In Oregon, only one contrast was significant but contrary to the hypothesis (β = .05, p < .001). Since the evidence to support H2e remained contradictory, we supported it with a time series visualization (Figure 3). The spike of novelty generated, after Pennsylvania was called (Period IV), signaling the victory of Biden, is noticeable in all search engines; at their peaks, "joe biden"'s novelty was highest in all cases (also in Frankfurt, Appendix S8), but we also noticed that its novelty quickly declined, and, at least, in Bing and DuckDuckGo, "donald trump"'s novelty increased after the spike (similar to previous values). The latter observation is consistent with the bias noted in Periods I to III (which rejected hypothesis H2e). Additionally, the spike of Biden's novelty in Period IV was strong enough to explain the two contrasts that did not support the hypothesis of a downward trend in novelty as time passes (H2d).

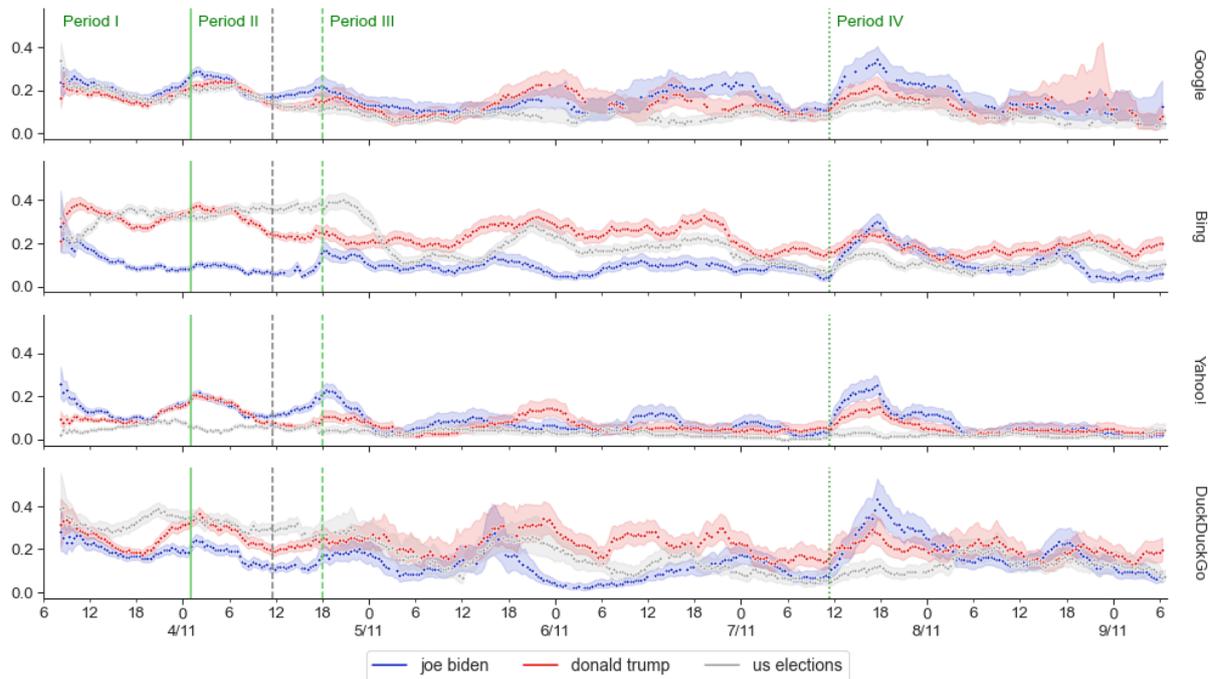

**Figure 3. Novelty of search results over time in Oregon.** The four plots present the rolled average novelty (of the last 6h, n=18) for each search engine (right label). The X-axis shows the day (major ticks) and hour (minor ticks) of the round in which the novelty was measured. The Y-axis shows the novelty truncated to .4 (maximum theoretical value 1.0). Each trace represents each of the query terms indicated on the legend. The green vertical lines divide each plot in four periods indicated in the label at the top. The gray dotted vertical line in Period II indicates the transition between collection A and B. Only the results collected in Oregon are shown. The bands indicate bootstrapped 95% confidence intervals.

## Discussion

Using the novelty of news results, we confirm that the US elections were widely covered by all search engines (H1a) except for Baidu, the only non-US search engine we included. The "coronavirus" query was the only other query that displayed similar novelty; in other cases, we do not find differences between the topical and stable queries, which highlights the tendency to neglect localized but topical and news-worthy happenings such as the Poland abortion protests and the Nagorno-Karabakh conflict (H1b).

Although we find several differences between the novelty displayed by each search engine (H2b), we only find partial support for the hypothesis that Google displays less novelty than Bing and Yahoo! (H2c)), as was the case for the main search results in the US elections 2016 due to spam control mechanisms (Metaxas & Pruksachatkun, 2017). Specifically, there is partial support for Bing, but not for Yahoo!. It is possible that these search engines have now implemented spam control mechanisms similar to those of Google (thus, changing the trends from 2016).

We find support for decreasing novelty as searches become more distant from the elections period (H2d). We find that there was a rebound in novelty for "joe biden" in Period IV due to the spike of news caused by the victory of Biden. Nevertheless, the spike did not compensate for the downward trend in all cases, as shown in the time series visualization (Figure 3).

We find differences in the novelty of the results concerning the election candidates in Periods I, II and III (H2e). It differs according to the search engine, with Bing and DuckDuckGo displaying more novelty for Trump, and Google and Yahoo! for Biden. This imbalance is particularly high for Bing, resulting in 3.24 times more unique news items for Trump. There are two possible reasons for this. First, Bing might replicate the imbalances of the news coverage that the candidates receive by news sources that are considered for the searches by this specific engine. Second, assuming that news sources were balanced, Bing's algorithm might have prioritized novel content related to Trump. In either case, the search engine shares some responsibility for such an imbalance that might be favoring the propagation of Trump's messages, including multiple claims of fraud (Benkler et al., 2020; Berlinski et al., 2021; Enders et al., 2021). Independently of the potential for spreading misinformation, for the period before polls closed (Period I), there are still potential undecided voters seeking last-minute information who might be exposed to a higher number of articles about Trump.

Crucially, for the period after Biden was declared the President Elect (Period IV), we predicted that there would be more novel news articles for "joe biden" (H2f), but we found the novelty to be resilient: after a spike in the novelty that favored Biden at the beginning of the period it shifted back to similar values of previous periods, for example, in Bing, the novelty favored Trump again after the spike. Another result that sets Bing apart is that it was the only search engine which consistently displayed more novelty for Biden in Frankfurt, contradicting H2a. Such attrition of novelty could be attributed to stronger spam mechanisms in Oregon, however, that explanation would make even more puzzling the higher novelty for "donald trump" in Oregon, as it would indicate that more content was blocked for "joe biden" for no apparent reason.

In line with previous research (Urman et al., 2021b), we find multiple differences between search engines. However, we observe some qualitative parallels between two search engine pairs in Oregon: Bing and DuckDuckGo, and Google and Yahoo!. First, Bing and DuckDuckGo showcased on average more novelty than Google and Yahoo!. Second, Bing and DuckDuckGo both displayed more novelty for Trump than for Biden, while Google and Yahoo! displayed close values of novelty for both candidates. DuckDuckGo explicitly acknowledges a relationship with Bing (DuckDuckGo, 2022) while Yahoo! has had partnerships, first with Bing (BBC, 2009) and then with Google until 2018 (Statt, 2015), though we were not able to verify if such partnerships still exist and if they extend to the news search. Shall this observation be correct, not only does Google capture 91.4% of the worldwide market, followed by Bing with 3.3% (Statcounter, 2022), but these two might influence the two here investigated alternatives, Yahoo! and DuckDuckGo, exacerbating a monopoly on the control of online information.

Aside from our findings, we present a series of methodological contributions. First, by establishing the volatility of news results during critical periods such as elections, we highlight the need for monitoring electoral processes longitudinally, where scattered snapshots might miss the big picture. Second, while other works have focused on the stability of search results (Metaxa et al., 2019), we introduced novelty, an indicator that measures how much new information is introduced. Third, we generalized the use of ranking bias used previously for political leaning (Kulshrestha et al., 2017, 2019) to our novelty measurement.

We list some limitations of our study. First, our results only cover one region in the US: Oregon. Instead of choosing two US regions, we decided to include Frankfurt as we were interested in international localization differences. Second, a list of three queries (of the same category) were assigned to each agent. Although the searches are synchronized across agents, the second query of the list is shifted 7 minutes (for each given round), and the third query, 14 minutes. Nevertheless, we argue that this should not affect the general patterns of the observed results as they are relatively small shifts. Third, we only include three queries per category. Fourth, we only analyze a small set of queries, but we point to potential spill overs that could emerge given the importance that search engines place on the recency of results. The observed imbalances could emerge (1) in other queries, for example, one could analyze if the novelty in the "us elections" is properly balanced between the candidates, and (2) in other sections of the search engines, for example, similar to the imbalances found for Google Top Stories section (Kawakami et al., 2020).

The present work also opens the door to future research regarding the visibility of candidates in news search, for example, the presence of each candidate in general queries (e.g., "us elections") or the news results that infiltrate the main search results. As the literature indicates, not only visibility, but also tonality, is important in terms of political choices (Hopmann et al., 2010), for which natural language processing techniques could be applied (Hutto & Gilbert, 2014). Finally, it is important to study the relation between novelty and information related to fraud claims and, in general, the presence of misinformation.

# Conclusion

The existent relation between news organizations and political campaigns continues its transformation, as digital intermediaries such as search engines leverage their influence to shape political communication. We started this investigation to learn how search engines process the quick turnover of news content generated during highly mediated political events such as the 2020 US elections. We argue that our metric, novelty, allows the investigation of coverage and visibility of topics in search engines, and we demonstrate differences across search engines, regions, and periods. We find an imbalance in novelty between the candidate queries, particularly large for Bing in Oregon. Contrary to the main web search, in which biases can be explained by the difficulty of balancing the enormous quantity of content available online, the number of available news articles is comparatively small and limited to a more defined set of sources that search engines already control for. Thus, search engines share a larger responsibility in providing a balanced coverage – either in their algorithms or in the criteria used in the selection of news sources. Such imbalances in novelty affect the visibility of political candidates in news search.

# References

Akaike, H. (1974). A new look at the statistical model identification. *IEEE Transactions on Automatic Control*, *19*(6), 716–723. https://doi.org/10.1109/TAC.1974.1100705

Astor, M. (2020, November 3). How to Follow the Election Results. *The New York Times*. https://www.nytimes.com/2020/11/03/us/politics/live-election-stream.html

Bates, D., Mächler, M., Bolker, B., & Walker, S. (2015). Fitting Linear Mixed-Effects Models Using lme4. *Journal of Statistical Software*, *67*, 1–48. https://doi.org/10.18637/jss.v067.i01

BBC. (2009, July 29). *Microsoft and Yahoo seal web deal*. http://news.bbc.co.uk/2/hi/business/8174763.stm

Benkler, Y., Tilton, C., Etling, B., Roberts, H., Clark, J., Faris, R., Kaiser, J., & Schmitt, C. (2020). *Mail-In Voter Fraud: Anatomy of a Disinformation Campaign* (SSRN Scholarly Paper No. 3703701). https://doi.org/10.2139/ssrn.3703701

Berlinski, N., Doyle, M., Guess, A. M., Levy, G., Lyons, B., Montgomery, J. M., Nyhan, B., & Reifler, J. (2021). The Effects of Unsubstantiated Claims of Voter Fraud on Confidence in Elections. *Journal of Experimental Political Science*, 1–16. https://doi.org/10.1017/XPS.2021.18

Boxell, L., Gentzkow, M., & Shapiro, J. M. (2020). *Cross-Country Trends in Affective Polarization* (Working Paper No. 26669; Working Paper Series). National Bureau of Economic Research. https://doi.org/10.3386/w26669

Box-Steffensmeier, J., Dillard, M., Kimball, D., & Massengill, W. (2015). The long and short of it: The unpredictability of late deciding voters. *Electoral Studies*, *39*, 181–194. https://doi.org/10.1016/j.electstud.2015.03.013

Callum Borchers. (2016, October 25). Yes, Donald Trump has been good for the media business. *Washington Post*. https://www.washingtonpost.com/news/the-fix/wp/2016/10/25/yes-donald-trump-has-been-good-for-the-media-business/

CNN. (2021). *Assault on Democracy: Paths to Insurrection*. https://www.cnn.com/interactive/2021/06/us/capitol-riot-paths-to-insurrection/

Confessore, N., & Yourish, K. (2016, March 15). $2 Billion Worth of Free Media for Donald Trump. *The New York Times*. https://www.nytimes.com/2016/03/16/upshot/measuring-donald-trumps-mammoth-advantage-in-free-media.html

Courtois, C., Slechten, L., & Coenen, L. (2018). Challenging Google Search filter bubbles in social and political information: Disconforming evidence from a digital methods case study. *Telematics and Informatics*, *35*(7), 2006–2015. https://doi.org/10.1016/j.tele.2018.07.004

Crowley, M. (2020, September 24). Trump Won't Commit to 'Peaceful' Post-Election Transfer of Power. *The New York Times*. https://www.nytimes.com/2020/09/23/us/politics/trump-power-transfer-2020-election.html

Dong, A., Zhang, R., Kolari, P., Bai, J., Diaz, F., Chang, Y., Zheng, Z., & Zha, H. (2010). Time is of the essence: Improving recency ranking using Twitter data. *Proceedings of the 19th International Conference on World Wide Web*, 331–340. https://doi.org/10.1145/1772690.1772725

DuckDuckGo. (2022). *Sources*. DuckDuckGo Help Pages. https://help.duckduckgo.com/duckduckgo-help-pages/results/sources/

Dutton, W. H., Reisdorf, B., Dubois, E., & Blank, G. (2017). *Social Shaping of the Politics of Internet Search and Networking: Moving Beyond Filter Bubbles, Echo Chambers, and Fake News* (SSRN Scholarly Paper ID 2944191). Social Science Research Network. https://doi.org/10.2139/ssrn.2944191

Edelman. (2021). *The 2021 Edelman Trust Barometer*. https://www.edelman.com/trust/2021-trust-barometer

Enders, A. M., Uscinski, J. E., Klofstad, C. A., Premaratne, K., Seelig, M. I., Wuchty, S., Murthi, M. N., & Funchion, J. R. (2021). The 2020 presidential election and beliefs about fraud: Continuity or change? *Electoral Studies*, *72*, 102366. https://doi.org/10.1016/j.electstud.2021.102366

Epstein, R., Robertson, R. E., Lazer, D., & Wilson, C. (2017). Suppressing the Search Engine Manipulation Effect (SEME). *Proceedings of the ACM on Human-Computer Interaction*, *1*(CSCW), 42:1-42:22. https://doi.org/10.1145/3134677

Fisher, M., Goddu, M. K., & Keil, F. C. (2015). Searching for explanations: How the Internet inflates estimates of internal knowledge. *Journal of Experimental Psychology: General*, *144*(3), 674–687. https://doi.org/10.1037/xge0000070

Fournier, P., Nadeau, R., Blais, A., Gidengil, E., & Nevitte, N. (2004). Time-of-voting decision and susceptibility to campaign effects. *Electoral Studies*, *23*(4), 661–681. https://doi.org/10.1016/j.electstud.2003.09.001

Ganguly, S., Kulshrestha, J., An, J., & Kwak, H. (2020). Empirical Evaluation of Three Common Assumptions in Building Political Media Bias Datasets. *Proceedings of the International AAAI Conference on Web and Social Media*, *14*, 939–943.

Gopoian, J. D., & Hadjiharalambous, S. (1994). Late-deciding voters in presidential elections. *Political Behavior*, *16*(1), 55–78. https://doi.org/10.1007/BF01541642

Hackett, R. A. (1984). Decline of a paradigm? Bias and objectivity in news media studies. *Critical Studies in Mass Communication*, *1*(3), 229–259. https://doi.org/10.1080/15295038409360036

Haim, M., Graefe, A., & Brosius, H.-B. (2018). Burst of the Filter Bubble? Effects of personalization on the diversity of Google News. *Digital Journalism*, *6*(3), 330–343. https://doi.org/10.1080/21670811.2017.1338145

Hamilton, J. T. (2011). All the news that's fit to sell. In *All the News That's Fit to Sell*. Princeton University Press.

Hofstetter, C. R., & Buss, T. F. (1980). Politics and last-minute political television. *Western Political Quarterly*, *33*(1), 24–37.

Hopmann, D. N., Vliegenthart, R., De Vreese, C., & Albæk, E. (2010). Effects of Election News Coverage: How Visibility and Tone Influence Party Choice. *Political Communication*, *27*(4), 389–405. https://doi.org/10.1080/10584609.2010.516798

Hutto, C., & Gilbert, E. (2014). VADER: A Parsimonious Rule-Based Model for Sentiment Analysis of Social Media Text. *Proceedings of the International AAAI Conference on Web and Social Media*, *8*(1), Article 1.

Jiang, M. (2014). Search Concentration, Bias, and Parochialism: A Comparative Study of Google, Baidu, and Jike's Search Results From China. *Journal of Communication*, *64*(6), 1088–1110. https://doi.org/10.1111/jcom.12126

Kaid, L. L., & Strömbäck, J. (2008). Election News Coverage Around the World: A Comparative Perspective. In *The Handbook of Election News Coverage Around the World*. Routledge.

Katz, E., & Lazarsfeld, P. F. (2017). *Personal Influence: The Part Played by People in the Flow of Mass Communications*. Routledge. https://doi.org/10.4324/9781315126234

Kawakami, A., Umarova, K., & Mustafaraj, E. (2020). The Media Coverage of the 2020 US Presidential Election Candidates through the Lens of Google's Top Stories. *Proceedings of the International AAAI Conference on Web and Social Media*, *14*, 868–877.

Keane, M. T., O'Brien, M., & Smyth, B. (2008). Are people biased in their use of search engines? *Communications of the ACM*, *51*(2), 49–52. https://doi.org/10.1145/1314215.1314224

Kliman-Silver, C., Hannak, A., Lazer, D., Wilson, C., & Mislove, A. (2015). Location, Location, Location: The Impact of Geolocation on Web Search Personalization. *Proceedings of the 2015 Internet Measurement Conference*, 121–127. https://doi.org/10.1145/2815675.2815714

Klinger, U., & Svensson, J. (2015). The emergence of network media logic in political communication: A theoretical approach. *New Media & Society*, *17*(8), 1241–1257. https://doi.org/10.1177/1461444814522952

Kommenda, N., Voce, A., Hulley-Jones, F., Leach, A., & Clarke, S. (2020, November 3). US election results 2020: Joe Biden defeats Donald Trump to win presidency. *The Guardian*. https://www.theguardian.com/us-news/ng-interactive/2020/nov/25/us-election-results-2020-joe-biden-defeats-donald-trump-to-win-presidency

Kreiss, D., & Mcgregor, S. C. (2018). Technology Firms Shape Political Communication: The Work of Microsoft, Facebook, Twitter, and Google With Campaigns During the 2016 U.S. Presidential Cycle. *Political Communication*, *35*(2), 155–177. https://doi.org/10.1080/10584609.2017.1364814

Kulshrestha, J., Eslami, M., Messias, J., Zafar, M. B., Ghosh, S., Gummadi, K. P., & Karahalios, K. (2017). Quantifying Search Bias: Investigating Sources of Bias for Political Searches in Social Media. *Proceedings of the 2017 ACM Conference on Computer Supported Cooperative Work and Social Computing*, 417–432. https://doi.org/10.1145/2998181.2998321

Kulshrestha, J., Eslami, M., Messias, J., Zafar, M. B., Ghosh, S., Gummadi, K. P., & Karahalios, K. (2019). Search bias quantification: Investigating political bias in social media and web search. *Information Retrieval Journal*, *22*(1), 188–227. https://doi.org/10.1007/s10791-018-9341-2

Lenth, R. V. (2021). *emmeans: Estimated Marginal Means, aka Least-Squares Means*. https://CRAN.R-project.org/package=emmeans

Maddens, B., Wauters, B., Noppe, J., & Fiers, S. (2006). Effects of Campaign Spending in an Open List PR System: The 2003 Legislative Elections in Flanders/Belgium. *West European Politics*, *29*(1), 161–168. https://doi.org/10.1080/01402380500389398

Makhortykh, M., Urman, A., & Ulloa, R. (2020). How search engines disseminate information about COVID-19 and why they should do better. *Harvard Kennedy School Misinformation Review*, *1*(COVID-19 and Misinformation). https://doi.org/10.37016/mr-2020-017

Metaxa, D., Park, J. S., Landay, J. A., & Hancock, J. (2019). Search Media and Elections: A Longitudinal Investigation of Political Search Results. *Proceedings of the ACM on Human-Computer Interaction*, *3*(CSCW), 129:1-129:17. https://doi.org/10.1145/3359231

Metaxas, P. T., & Mustafaraj, E. (2009). *The Battle for the 2008 US Congressional Elections on the Web*. https://repository.wellesley.edu/islandora/object/ir%3A314/

Metaxas, P. T., & Pruksachatkun, Y. (2017). *Manipulation of Search Engine Results during the 2016 US Congressional Elections*. https://repository.wellesley.edu/islandora/object/ir%3A264/

Minnite, L. C. (2011). The Myth of Voter Fraud. In *The Myth of Voter Fraud*. Cornell University Press. https://doi.org/10.7591/9780801459061

Mittelstadt, B. (2016). Automation, Algorithms, and Politics| Auditing for Transparency in Content Personalization Systems. *International Journal of Communication*, *10*(0), 12.

Muñoz-Torres, J. R. (2012). Truth and Objectivity in Journalism. *Journalism Studies*, *13*(4), 566–582. https://doi.org/10.1080/1461670X.2012.662401

Nechushtai, E., & Lewis, S. C. (2019). What kind of news gatekeepers do we want machines to be? Filter bubbles, fragmentation, and the normative dimensions of algorithmic recommendations. *Computers in Human Behavior*, *90*, 298–307. https://doi.org/10.1016/j.chb.2018.07.043

Nielsen. (2020, November 4). *Media Advisory: 2020 Election Coverage Draws 56.9 Million Viewers During Prime*. https://www.nielsen.com/us/en/press-releases/2020/media-advisory-2020-election-draws-56-9-million-viewers-during-prime

Nielsen, R. K., & Ganter, S. A. (2018). Dealing with digital intermediaries: A case study of the relations between publishers and platforms. *New Media & Society*, *20*(4), 1600–1617. https://doi.org/10.1177/1461444817701318

Pan, B., Hembrooke, H., Joachims, T., Lorigo, L., Gay, G., & Granka, L. (2007). In Google We Trust: Users' Decisions on Rank, Position, and Relevance. *Journal of Computer-Mediated Communication*, *12*(3), 801–823. https://doi.org/10.1111/j.1083-6101.2007.00351.x

Pariser, E. (2011). *The Filter Bubble: How the New Personalized Web Is Changing What We Read and How We Think*. Penguin.

Patterson, T. E. (2013). *Informing the news: The need for knowledge-based journalism*. Vintage.

Pradel, F. (2021). Biased Representation of Politicians in Google and Wikipedia Search? The Joint Effect of Party Identity, Gender Identity and Elections. *Political Communication*, *38*(4), 447–478. https://doi.org/10.1080/10584609.2020.1793846

Puschmann, C. (2019). Beyond the Bubble: Assessing the Diversity of Political Search Results. *Digital Journalism*, *7*(6), 824–843. https://doi.org/10.1080/21670811.2018.1539626

Reuning, K., & Dietrich, N. (2019). Media Coverage, Public Interest, and Support in the 2016 Republican Invisible Primary. *Perspectives on Politics*, *17*(2), 326–339. https://doi.org/10.1017/S1537592718003274

Robertson, R. E., Jiang, S., Joseph, K., Friedland, L., Lazer, D., & Wilson, C. (2018). Auditing Partisan Audience Bias within Google Search. *Proceedings of the ACM on Human-Computer Interaction*, *2*(CSCW), 148:1-148:22. https://doi.org/10.1145/3274417

Samuel-Azran, T., Yarchi, M., & Hayat, T. Z. (2022). Less critical and less informed: Undecided voters' media (dis)engagement during Israel's April 2019 elections. *Information, Communication & Society*, *25*(12), 1752–1768. https://doi.org/10.1080/1369118X.2021.1883706

Schaul, K., Rabinowitz, K., & Mel, T. (2020, November 5). 2020 turnout is the highest in over a century. *Washington Post*. https://www.washingtonpost.com/graphics/2020/elections/voter-turnout/

Schill, D., & Kirk, R. (2017). Angry, Passionate, and Divided: Undecided Voters and the 2016 Presidential Election. *American Behavioral Scientist*, *61*(9), 1056–1076. https://doi.org/10.1177/0002764217709040

Schmitt-Beck, R. (2003). Mass Communication, Personal Communication and Vote Choice: The Filter Hypothesis of Media Influence in Comparative Perspective. *British Journal of Political Science*, *33*(2), 233–259. https://doi.org/10.1017/S0007123403000103

Schmitt-Beck, R., & Partheymüller, J. (2012). Why Voters Decide Late: A Simultaneous Test of Old and New Hypotheses at the 2005 and 2009 German Federal Elections. *German Politics*, *21*(3), 299–316. https://doi.org/10.1080/09644008.2012.716042

Schultheiß, S., Sünkler, S., & Lewandowski, D. (2018). We still trust in Google, but less than 10 years ago: An eye-tracking study. *Information Research*, *23*, paper 799.

Shafer, J. (2016, November). *How Trump Took Over the Media By Fighting It*. POLITICO Magazine. https://www.politico.com/magazine/story/2016/11/2016-election-trump-media-takeover-coverage-214419

Silver, N. (2017, January 23). The Invisible Undecided Voter. *FiveThirtyEight*. https://fivethirtyeight.com/features/the-invisible-undecided-voter/

Statcounter. (2022). *Search Engine Market Share Worldwide*. StatCounter Global Stats. https://gs.statcounter.com/search-engine-market-share

Statt, N. (2015, October 20). *Yahoo enters deal to display Google search results*. The Verge. https://www.theverge.com/2015/10/20/9577519/yahoo-google-search-deal

Steiner, M., Magin, M., Stark, B., & Geiß, S. (2020). Seek and you shall find? A content analysis on the diversity of five search engines' results on political queries. *Information, Communication & Society*, *0*(0), 1–25. https://doi.org/10.1080/1369118X.2020.1776367

Stier, S., Mangold, F., Scharkow, M., & Breuer, J. (2022). Post Post-Broadcast Democracy? News Exposure in the Age of Online Intermediaries. *American Political Science Review*, *116*(2), 768–774. https://doi.org/10.1017/S0003055421001222

Trielli, D., & Diakopoulos, N. (2019). Search as news curator: The role of Google in shaping attention to news information. *Proceedings of the 2019 CHI Conference on Human Factors in Computing Systems*, 1–15.

Ulloa, R., & Kacperski, C. S. (2022). *Search engine effects on news consumption: Ranking and representativeness outweigh familiarity in news selection* (arXiv:2206.08578). arXiv. https://doi.org/10.48550/arXiv.2206.08578

Ulloa, R., Makhortykh, M., & Urman, A. (2021). Algorithm Auditing at a Large-Scale: Insights from Search Engine Audits. *ArXiv:2106.05831 [Cs]*. http://arxiv.org/abs/2106.05831

Unkel, J., & Haim, M. (2019). Googling Politics: Parties, Sources, and Issue Ownerships on Google in the 2017 German Federal Election Campaign. *Social Science Computer Review*, 0894439319881634. https://doi.org/10.1177/0894439319881634

Urman, A., & Makhortykh, M. (2021). You Are How (and Where) You Search? Comparative Analysis of Web Search Behaviour Using Web Tracking Data. *ArXiv:2105.04961 [Cs]*. http://arxiv.org/abs/2105.04961

Urman, A., Makhortykh, M., & Ulloa, R. (2021a). Auditing Source Diversity Bias in Video Search Results Using Virtual Agents. *Companion Proceedings of the Web Conference 2021*, 232–236. https://doi.org/10.1145/3442442.3452306

Urman, A., Makhortykh, M., & Ulloa, R. (2021b). The Matter of Chance: Auditing Web Search Results Related to the 2020 U.S. Presidential Primary Elections Across Six Search Engines. *Social Science Computer Review*, 08944393211006863. https://doi.org/10.1177/08944393211006863

Van Aelst, P., Strömbäck, J., Aalberg, T., Esser, F., de Vreese, C., Matthes, J., Hopmann, D., Salgado, S., Hubé, N., Stępińska, A., Papathanassopoulos, S., Berganza, R., Legnante, G., Reinemann, C., Sheafer, T., & Stanyer, J. (2017). Political communication in a high-choice media environment: A challenge for democracy? *Annals of the International Communication Association*, *41*(1), 3–27. https://doi.org/10.1080/23808985.2017.1288551

van Erkel, P. F. A., Van Aelst, P., & Thijssen, P. (2020). Does media attention lead to personal electoral success? Differences in long and short campaign media effects for top and ordinary political candidates. *Acta Politica*, *55*(2), 156–174. https://doi.org/10.1057/s41269-018-0109-x

Ward, A. F. (2021). People mistake the internet's knowledge for their own. *Proceedings of the National Academy of Sciences*, *118*(43). https://doi.org/10.1073/pnas.2105061118

Willocq, S. (2019). Explaining Time of Vote Decision: The Socio-Structural, Attitudinal, and Contextual Determinants of Late Deciding. *Political Studies Review*, *17*(1), 53–64. https://doi.org/10.1177/1478929917748484

Yarchi, M., Wolfsfeld, G., & Samuel-Azran, T. (2021). Not all undecided voters are alike: Evidence from an Israeli election. *Government Information Quarterly*, *38*(4), 101598. https://doi.org/10.1016/j.giq.2021.101598

Zweig, K. (2017, April 7). *Watching the watchers: Epstein and Robertson's „Search Engine Manipulation Effect"*. AlgorithmWatch. https://algorithmwatch.org/en/watching-the-watchers-epstein-and-robertsons-search-engine-manipulation-effect/